\documentclass[11pt,letterpaper]{article}
\usepackage[hang, small, bf, margin=5pt, tableposition=top]{caption}
\usepackage{sectsty}
\sectionfont{\normalsize}
\usepackage{multicol}
\usepackage{latexsym}
\usepackage{amsmath}
\usepackage{latexsym,amssymb}
\usepackage{graphicx}
\usepackage{bm}
\usepackage{enumerate}
\usepackage{dsfont}
\usepackage{bbm}
\usepackage{setspace}
\usepackage{framed}
\newcommand{\romansubs}{\renewcommand{\theequation}{\theparentequation \roman{equation}}}
\newcommand{\HRule}{\rule{\linewidth}{0.5mm}}

\newcommand{\tinymath}[1]{\mbox{\tiny{#1}}}

\newcommand{\athree}{\mathcal{A}_{\tinymath{III}}}
\newcommand{\aone}{\mathcal{A}_{\tinymath{I}}}
\newcommand{\atwo}{\mathcal{A}_{\tinymath{II}}}
\newcommand{\ethree}{\mathcal{E}_{\tinymath{III}}}
\newcommand{\eone}{\mathcal{E}_{\tinymath{I}}}
\newcommand{\etwo}{\mathcal{E}_{\tinymath{II}}}

\newcommand{\ppar}{p_{\mbox{\tiny{$\parallel$}}}}

\newcommand{\pperp}{p_{\mbox{\tiny{$\perp$}}}}
\newcommand{\af}{{\mbox{\tiny{af}}}}
\newcommand{\detE}{\mbox{det}(E)}
\newcommand{\sqrtdetE}{\sqrt{\left|\ethree\right|} \cdot \left|\sin(\theta)\right|\cdot \mathsf{E}}
\newcommand{\deltatwo}{\delta_{\mbox{\tiny{II}}}}
\newcommand{\grr}{1+\left[\partial_{r}P_{\pm}(r)\right]^{2}}
\newcommand{\gxx}{1+\left[\partial_{x}Q(x)\right]^{2}}
\newcommand{\aphi}{\mathcal{A}_{\phi}}
\newcommand{\ephi}{\mathcal{E}^{\phi}}
\newcommand{\qzero}{Q_{0}}
\newcommand{\mina}{a_{\mbox{\tiny{min}}}}
\newcommand{\PI}[2]{\underset{\mbox{\hspace{-0.15cm}{\tiny (#1)}}}{\Pi_{#2}}}
\newcommand{\qe}[4]{\frac{E_{#1}^{\;#2}E_{#3}^{\;#4}}{\mbox{det}(E)}}

\def \u1 {\textrm{U(1)}}

\begin{document}
\title{\bf{\Large Anisotropic Structures and Wormholes with Loop Quantum Gravity Holonomy Corrections}}
\author{{{\small Andrew DeBenedictis \footnote{adebened@sfu.ca}}} \\
\it{\small Department of Physics,} \\
\it{\small Simon Fraser University}\\
{\small {and}} \\
\it{\small The Pacific Institute for the Mathematical Sciences,} \\[0.3cm]
\it{\small Burnaby, British Columbia, V5A 1S6, Canada}\\ \HRule 
}
\date{}
\maketitle

\vspace{-0.5cm}
\begin{abstract}
Anisotropic spherically symmetric systems are studied in the connection and densitized triad variables used in loop quantum gravity. The material source is an anisotropic fluid, which is arguably the most commonly used source term in anisotropic studies within general relativity. The gravitational+anisotropic fluid constraints are derived and analyzed and then quantum gravity inspired holonomy replacements are performed. The quantum properties of the fluid are dictated by the modified constraint equations. Particular attention is paid to wormhole throats, as they provide a simplistic model of the structures thought to be ubiquitous  in the quantum gravity space-time foam at high energy scales. In comparison to the purely classical theory, the quantum corrections act to increase the energy density of the fluid, which indicates that they may lessen the energy condition violation present in the classical theory. Related to this, in principle it would be possible to have scenarios where the classical solution yields everywhere negative (with a zero at the throat) fluid energy density but the corresponding quantum corrected theory possesses only small regions of negative energy density or even everywhere non-negative energy density.
\end{abstract}
\rule{\linewidth}{0.25mm}
\vspace{-0.5mm}
{\noindent{\small{PACS numbers: 04.20.Gz, 04.60.Pp, 04.60.-m}}}\\

\begin{center}{\section{\hspace{-0.35cm}:\hspace{0.33cm}\small INTRODUCTION}}\end{center}
In classical general relativity anisotropy has often played an important role from models of stellar structure and cosmologies to more exotic solutions \cite{ref:exactsols1}-\cite{ref:sasathesis}. In the category of exotic solutions exhibiting spherical symmetry, anisotropy is almost always present in systems such as wormholes, gravastars, and even certain black hole models forming from gravitational collapse or singularity-free models. (See, for example, \cite{ref:kimcqworm} - \cite{ref:singfreesaibal} and references therein.) These sorts of anisotropic solutions usually involve strong gravity effects, and in some cases it may be argued that they are in the realm where effects from a quantum theory of gravitation should be considered. For example, the vicinity of the classical singularity formed during the gravitational collapse is almost certainly a domain where quantum gravity effects are expected to apply. More recently, Bojowald, Paily, Reyes and Tibrewala have performed a thorough, in-depth study of midisuperspace models containing horizons with scalar fields \cite{ref:Bojetal} in order to study quantum effects in these space-times. 

\begin{figure}[h!t]
\begin{framed}
\begin{center}
\vspace{0.5cm}
\includegraphics[bb=0 0 892 285, scale=0.33, clip, keepaspectratio=true]
{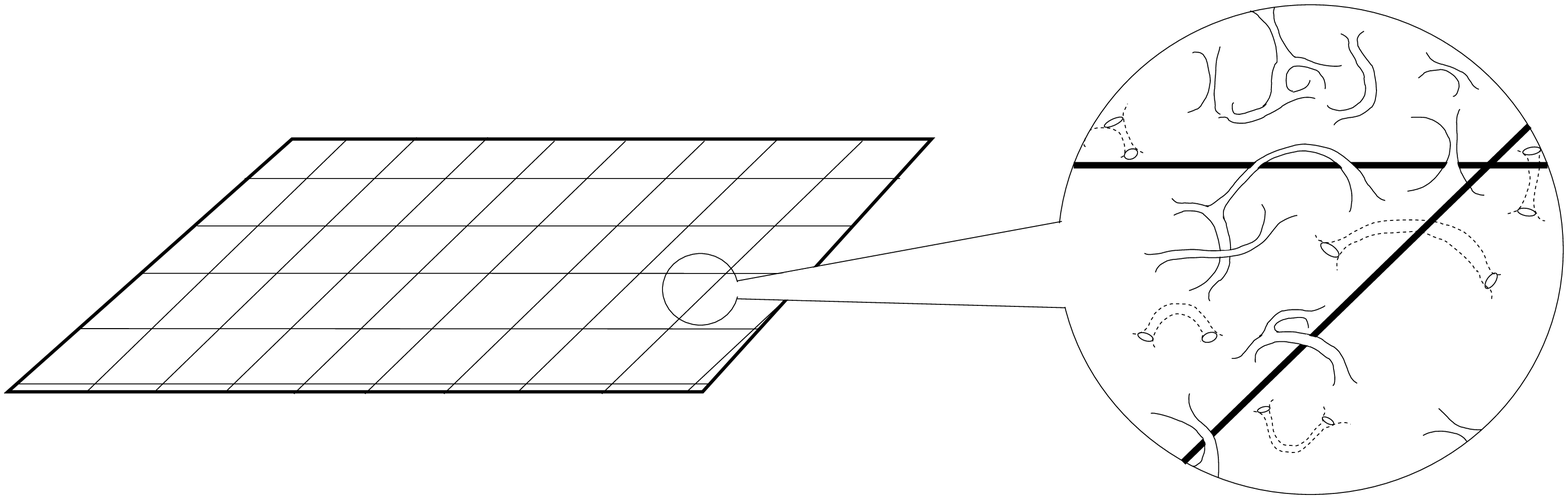}
\end{center}\vspace{-0.1cm}
\caption{{A possible picture for the space-time foam. Space-time that seems smooth on large scales (left) may actually be endowed with fluctuating topologies (represented by handles on the right) due to quantum gravity effects. One of the simplest models for such a handle is the wormhole.}}%
\label{fig:foam}%
\end{framed}
\end{figure}

In the arena of non-trivial topologies, a common rationale for studying wormholes is that they may provide a simple model for the space-time foam, a potential model for the vacuum in theories of quantum gravity, where not only the geometry but also the topology of space-time may fluctuate due to quantum effects, \cite{ref:Wheeler}, \cite{ref:geons}, as illustrated in figure \ref{fig:foam}. It has been argued that systems of wormholes may provide good models for such a foam \cite{ref:foamworm}, \cite{ref:foamworm2}, \cite{ref:foamworm3}. Therefore, in the vein of wormhole-like structures, there is good reason to study these systems utilizing the variables of loop quantum gravity due to their relevance in topology change which may exist in the quantum gravity realm. In the context of black hole physics, topologically non-trivial horizons have already been studied within loop quantum gravity \cite{ref:kbd}-\cite{ref:bammod}.

\begin{figure}[h!t]
\begin{framed}
\begin{center}
\vspace{0.5cm}
\includegraphics[bb=0 0 900 410, scale=0.33, clip, keepaspectratio=true]
{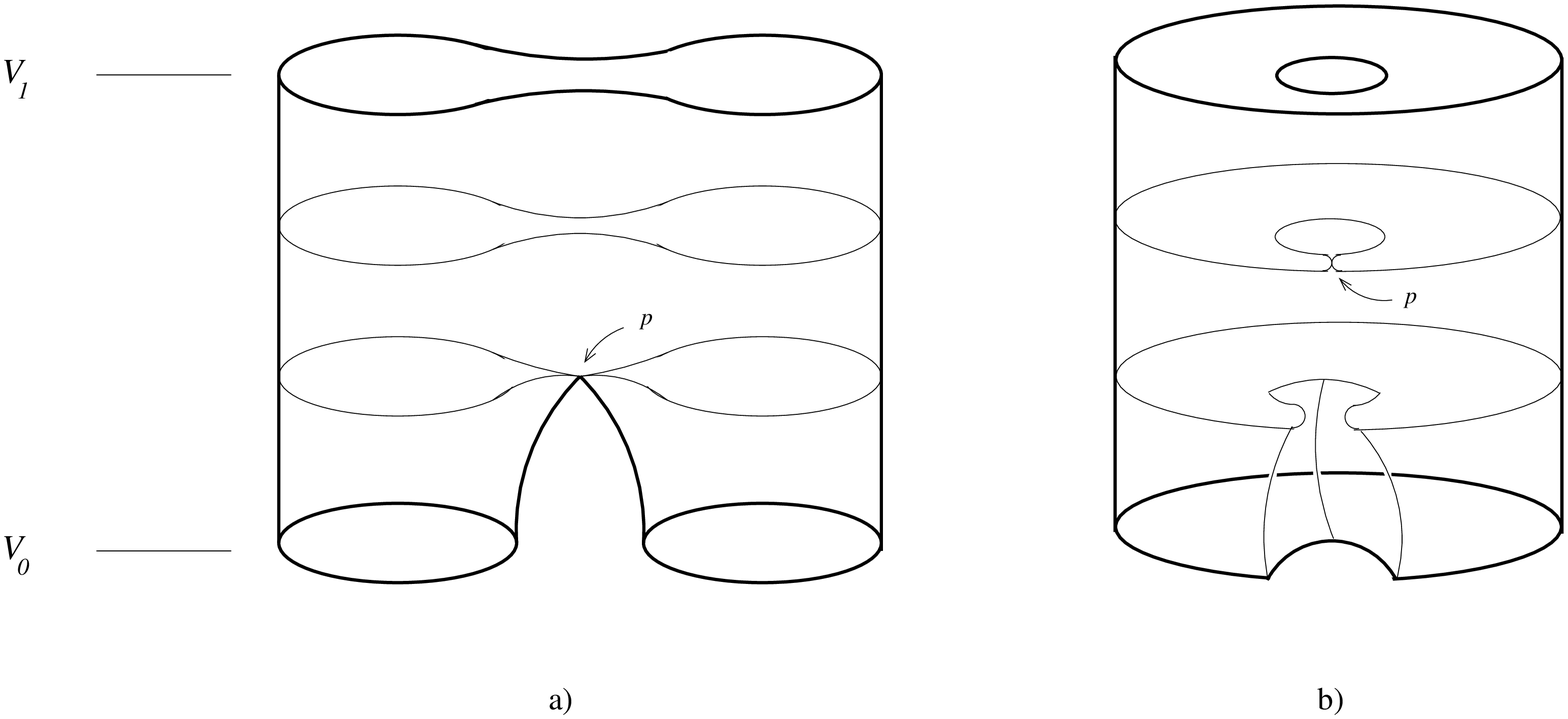}
\end{center}\vspace{-0.1cm}
\caption{{Schematics of topology changing space-times via wormhole
formation. Figure a) represents topology change via the formation of an
inter-universe wormhole. Figure b) represents topology change via the
formation of an intra-universe wormhole. The points $p$ represent the critical
point of the topology change, where quantum gravitation effects are expected to become important. This sort of topology change is thought to be ubiquitous at the Planck scale.}}%
\label{fig:tc1}%
\end{framed}
\end{figure}

There have been a number works produced studying the possibility of a change in the topology of space-time at the classical and semi-classical level \cite{ref:reinhart} - \cite{ref:phantomstar}. It is known that if one allows for the possibility of topology change in space-time, there are arguments that if $V_{0}$ and
$V_{1}$ are distinct compact 3-manifolds, there will exist a space-time
whose boundary is comprised of the disjoint union of $V_{0}$ and
$V_{1}$ \cite{ref:reinhart} (see figure \ref{fig:tc1} for
reference.) Regarding this, Geroch's theorem states that if a segment of space-time has $V_{0}$
and $V_{1}$ as boundaries possessing differing topology, then a singularity or
closed time-like curves must exist somewhere on the manifold \cite{ref:geroch}. It has been convincingly argued that these singularities that arise in certain topology changing space-times are extremely mild \cite{ref:horo} in the sense that the tetrad
becomes degenerate (introducing metric degeneracy) but the resulting curvature remains well defined. The loop quantum gravity approach utilized here relies on \emph{densitized tetrads} and SU(2) connections, and it is known that solutions with classically degenerate metrics can yield finite equations when using these new Ashtekar variables \cite{ref:rovreport}. Therefore, even before implementing quantum corrections it is worthwhile using these alternate variables as they are better suited to study systems with non-trivial topology. In fact, it is possible that degenerate tetrads may play an important role in quantum gravity \cite{ref:tseyt}, \cite{ref:wittentop}.

The cases for topology change or the existence of a space-time foam are still open. It is not yet completely clear if either of these are indeed possible or not. However, they are actually related issues as the existence of a space-time foam as a sea of fluctuating topologies is directly linked to the possibility of topology change. That is, if this picture of the space-time foam is correct, then topology change is allowed, at least in the quantum gravity realm. Although we also cannot answer this question definitively here, we do find that the existence of non-trivial topologies is less exotic when loop quantum corrections are present in the equations of motion, and hence this may be an indication that in the full loop-quantum gravity picture topology change would be at least more probable than in the classical regime. In Wheeler-DeWitt theory, a study has also been performed by Visser \cite{ref:viswdw} to determine whether the existence of non-trivial topologies is favored or not in that theory. 

As well, the existence of wormholes in gravity theories which deviate from general relativity in the high curvature (high energy) realm are of interest to the wormhole community. Of particular recent interest has been $f(R)$ theory, where higher curvature scalar terms may appear in the gravitational Lagrangian. (See, for example, \cite{ref:fofr1}.) Wormholes have been studied in the $f(R)$ theory \cite{ref:hochworm}, \cite{ref:lobofrworm}, \cite{ref:bronwh}, \cite{ref:fofrlast}. The study in this article can be seen as another extension of general relativity to the high energy realm. Specifically related to discerning effects from quantum gravity, some studies of wormholes within the paradigm of Wheeler-DeWitt theory have been carried out in \cite{ref:kimcqworm}, \cite{ref:remowdw}. As well, many authors have used classical higher derivative theories of gravity to capture some possible effects of high energy corrections to the general relativity solutions \cite{ref:fukuhd}, \cite{ref:ghorsom}, \cite{ref:fureydeb}, some of which fall in the $f(R)$ category mentioned above. Others still have approached the problem by considering semi-classical gravity where the non-trivial topologies are supported by quantum matter \cite{ref:sushsemi} - \cite{ref:kuhsemiclass}.

Given that the above solutions are necessarily anisotropic, we aim to study here inhomogeneous anisotropic systems. As most of the studies of such solutions have to date been classical, we also add some low-order loop quantum gravity holonomy corrections. We use a midi-superspace approach where we freeze the symmetry a priori to be spherical symmetry and then apply an ``effective quantization'' by replacing the configuration variable (the SU(2) connection) with a function incorporating some holonomy corrections from the quantum theory. Although we analyze anisotropic structures in general, we will pay special attention to the wormhole solutions, given the above arguments for their possible importance in quantum gravity. 

Regarding the possible detection of quantum gravity effects at the macroscopic level, it was suggested in Ref. \cite{ref:Morris} that one could imagine an extremely advanced civilization \cite{ref:Lemos:2003jb} pulling a wormhole from this submicroscopic space-time quantum foam and enlarging it to macroscopic dimensions. However, in a more plausible scenario, the possibility that inflation might provide a natural mechanism for the enlargement of such wormholes to macroscopic size was explored \cite{ref:Roman:1992xj}. Some authors have also investigated the effects of such a foamy space on the cosmological constant. One example is the Coleman mechanism, where micro-wormhole contributions suppress the cosmological constant, explaining its small observed value \cite{ref:Coleman}. However, even in the absence of detectable effects, predictions of a quantum gravity theory are important for understanding physics at the Planck scale \cite{ref:physphil}.

\vspace{0.0cm}
\begin{center}{\section{\hspace{-0.35cm}:\hspace{0.33cm}\small{SPHERICALLY SYMMETRIC ANISOTROPIC STRUCTURES AND QUANTUM CORRECTIONS}}}
\end{center}\vspace{-0.0cm}
The basic variables in loop quantum gravity (fixed to the time gauge) are the SU(2) connection as the configuration variable and the densitized triad as the conjugate momentum variable. In terms of more traditional variables in the 3+1 space-time decomposition, the densitized triad, $E_{i}^{\;a}$, is directly related to the spatial three-metric, $q_{ab}$, via
\begin{equation}
E_{i}^{\;a}E_{j}^{\;b}\,\delta^{ij}=\mbox{det}(q)\, q^{ab}\,. \label{eq:Eqrel}
\end{equation}
In this work we adopt the convention that indices $a$, $b$, $c$, etc. denote spatial components whereas indices $i$, $j$, $k$, etc. are internal indices coupling to the su(2) algebra.

The SU(2) connection encodes information about the extrinsic curvature of the spatial 3-surface via
\begin{equation}
A^{i}_{\;a}:=\Gamma^{i}_{\;a}+\gamma K^{i}_{\;a}, \label{eq:biconnection}
\end{equation} 
where $\gamma$ is the Barbero-Immirzi parameter. Here, $\Gamma^{i}_{\;a}$ is the spin connection whose associated derivative annihilates the regular (i.e. non-densitized) inverse-triad, $e^{i}_{\;b}$:
\begin{equation}
 \partial_{[a}e^{i}_{\;b]}+\epsilon^{i}_{\;jk}\Gamma^{j}_{\;[a}e^{k}_{\;b]}=0\,, \label{eq:cartan}
\end{equation}
and $K^{i}_{\;a}:= \frac{1}{\sqrt{\det(E)}}\delta^{ij} E_{j}^{\;b}K_{ab}\,$ ($K_{ab}$ being the usual extrinsic curvature of the 3-surface). $\Gamma^{i}_{\;a}$ may be explicitly solved for via inversion of (\ref{eq:cartan}) as
\begin{equation}
 \Gamma^{i}_{\;a}=-\frac{1}{2}\epsilon^{ij}_{\;\;\;k}e^{\;\;b}_{j}\left[\partial_{a}e^{k}_{\;b}-\partial_{b}e^{k}_{\;a} +\delta^{kl}\delta_{mn}e_{l}^{\;c}\,e^{m}_{\;\,a}\,\partial_{b}e^{n}_{\;c}\right]\,. \label{eq:Gam}
\end{equation}

Writing the Einstein-Hilbert action in the 3+1 form of the ADM formalism in terms of $A$ and $E$, and varying with respect to the shift-vector, $N^{a}$, and the lapse function, $N$, yields the gravitational vector constraint and scalar constraint respectively:
\begin{subequations}
\romansubs
\begin{align}
{V}_{a}^{\mbox{\tiny{(grav)}}} =& {E^{\;b}_{i}F^{i}_{\;ab}}=0, \label{eq:vecconst} \\
{\mathcal{S}}^{\mbox{\tiny{(grav)}}}=&\left\{\frac{{E}^{\;a}_{i}{E}^{\;b}_{j}}{\sqrt{\det({E})}}  \left[\epsilon^{ij}_{\;\;\;k}{F}^{k}_{\;ab} -2(1+\gamma^{2}) {K^{i}_{\;[a}K^{j}_{\;b]}}\right] +2{\Lambda}\sqrt{\det({E})}  \right\} =0, \label{eq:scaleconst}
\end{align}
\end{subequations}
where we have included the cosmological constant ($\Lambda$) term for generality (which we set to zero in the subsequent analysis). The field-strength is given by $F^{i}_{\;ab}:= \partial_{a}A^{i}_{\;b} -\partial_{b} A^{i}_{\;a} +\epsilon^{i}_{\;\;jk}A^{j}_{\;a}A^{k}_{\;b}$. To this set one adds the Gauss constraint in order to eliminate redundant degrees of freedom from internal rotations:
\begin{equation}
 {G}_{i}^{\mbox{\tiny{(grav)}}}:=\partial_{a}E^{\;a}_{i} +\epsilon_{ij}^{\;\;\;k}A^{j}_{\;a}E^{\;a}_{k}=:{D_{a}E^{\;a}_{i}} =0. \label{eq:gaussconst}
\end{equation}

The formal quantization occurs by replacing the ``classical functions'' $A^{i}_{\;a}$ and $E_{i}^{\;a}$ in (\ref{eq:vecconst},ii) and (\ref{eq:gaussconst}) with operators which obey the non-canonical algebra of loop quantum gravity. The resulting constraint operators then serve to constrain a ray in the Hilbert space with an appropriate inner-product \cite{ref:AQG}, \cite{ref:RQG}, \cite{ref:TQG}.

As dealing within the full quantum theory is technically difficult, we will deal with an effective quantum theory such as that which is often used in studies of loop quantum cosmology and the study of vacuum black hole interiors. That is, the connection shall be replaced with functions of $A$ which can more easily be represented by holonomies. Details on this procedure shall be provided later. Specifically, it has been shown that, within this approximation, the true singularities present in the classical theory are alleviated in the corresponding quantum theory in both black hole interiors as well as in cosmological models. (See \cite{ref:bojlrr}, \cite{ref:adbhsingrev} and references therein for details). As well, much progress has been made regarding the partial quantization of spherically symmetric inhomogeneous pure gravity in the interesting works on the exterior Schwarzschild space-time \cite{ref:kuchar}, \cite{ref:campschwarz}, \cite{ref:campschwarz2}, \cite{ref:gambpul}. 

The systems studied here are necessarily inhomogeneous, anisotropic, and non-vacuum, which leads to complications not present in simpler systems. To begin with, we will require an appropriate inhomogeneous and anisotropic matter contribution to the equations (\ref{eq:vecconst},ii) and (\ref{eq:gaussconst}). For this we choose an anisotropic fluid whose four-dimensional stress-energy tensor is given by:
\begin{equation}
T_{\mu\nu}=(\rho + \pperp)u_{\mu}u_{\nu} + \pperp\,g_{\mu\nu} + (\ppar - \pperp)s_{\mu}s_{\nu}\,. \label{eq:anisoT}
\end{equation}
Here $\rho$, $\pperp$ and $\ppar$ are the energy density, perpendicular (to the inhomogeneous direction) pressure and parallel pressure respectively as measured in the fluid element's rest frame. The vector $u^{\mu}$ is the fluid 4-velocity and $s^{\mu}$ is a space-like vector orthogonal to $u^{\mu}$. On-shell these vectors satisfy:
\begin{equation}
 u^{\mu}u_{\mu}=-1,\;\;\; s^{\mu}s_{\mu}=+1,\;\;\; u^{\mu}s_{\mu}=0\,. \label{eq:fluidvecs}
\end{equation}

The fluid is particularly useful in anisotropic studies as it provides one of the most general matter models allowed, yet still respecting the symmetry constraints. Essentially, for anisotropic structures in spherical symmetry we require a stress-energy tensor of Segre characteristic $[1,\,1,\,(1,\,1)]$, and the spherically symmetric anisotropic fluid is one of the most general matter models which provide this. It may also be noted that the material (\ref{eq:anisoT}) is also capable of accommodated Segre characteristic $[1,\,(1,\,1,\,1)]$  (i.e. the perfect fluid) in the special limit $\pperp=\ppar$ and can therefore be seen as a generalization on perfect fluids. A possibly more fundamental material, such as the scalar field, under certain assumptions on the field configuration which are compatible with spherical symmetry, is algebraically similar (in its stress-energy tensor) to the anisotropic or isotropic fluid \cite{ref:dasscalars}, and therefore by choosing a fluid source we potentially cover these scenarios as well. Finally, using the anisotropic fluid will simplify comparisons with purely classical solutions which most often use a fluid material as their source. 

Hamiltonians for a dust and an isotropic perfect fluid have already been constructed in the literature \cite{ref:taubham}-\cite{ref:brdham} and, as is usual in fluid mechanics, variables more appropriate for thermodynamic studies are usually chosen. However, here we choose to work in the original fluid variables of (\ref{eq:anisoT}) as this system is more perspicuous and has immediate physical interpretation in these variables. The price to pay is that certain relationships between thermodynamic quantities (such as temperature, entropy, etc.) are not enforced from the variational principle in this scheme, but these are usually not of primary interest in gravitational studies. For an in-depth treatment of more fundamental field couplings to loop quantum gravity see \cite{ref:ashmatter}, \cite{ref:sahlmatter}.

As a fluid of this type is considered an effective matter model, there seems to be no Lagrangian in the literature describing the anisotropic fluid. Therefore we adopt here a pragmatic strategy often used in field theories \cite{ref:goldstein}. Specifically, we will construct a Lagrangian density, $\mathcal{L}_{\af}$, which produces the desired equations of motion, in this case the stress-energy tensor (\ref{eq:anisoT}). After the derivation, the conservation law (which results from having a true scalar Lagrangian and the gravitational field equations) supplemented with the enforcement of the on-shell conditions (\ref{eq:fluidvecs}), which are also consequences of the variational principle in this scheme, will guarantee that the fluid equations of motion are obeyed. We propose the following matter action which produces the desired result:
\begin{align}
 I_{\af}=&\int_{M^{4}} \mathcal{L}_{\af}\,d^{4}x \nonumber \\ 
=&8\pi\int_{M^{4}} \sqrt{-g} \,\left\{ \Theta \left[u_{\alpha}u_{\beta}\,g^{\alpha\beta} +1 \right] -2\pperp + \Delta\left[s_{\alpha}s_{\beta}\,g^{\alpha\beta} -1 \right]\right\}\,d^{4}x\,, \label{eq:fluidact}
\end{align}
where we define $\Theta:=(\rho + \pperp)$ and $\Delta:=(\ppar-\pperp)$, which must be viewed as independent variables in their own right (the quantity $\Theta$ can be viewed as the particle number times the enthalpy). The quantities $\rho$ and $\ppar$ are true scalars as can be easily checked using (\ref{eq:anisoT}) and (\ref{eq:fluidvecs}), via: $T^{\mu\nu}u_{\mu}u_{\nu} = \rho$ and $T^{\mu\nu}s_{\mu}s_{\nu} = \ppar$. Now, having established that $\rho$ and $\ppar$ are scalars, the trace $T^{\mu}_{\;\,\mu}= -\rho +2\pperp + \ppar$ establishes that $\pperp$ must also be a scalar quantity (and hence so are $\Theta$ and $\Delta$). Note that with this particular choice of action the fluid variables $\Theta$ and $\Delta$ appear simply as Lagrange multiplier fields which give rise to the set of constraints (\ref{eq:fluidvecs}), and their dynamics are not completely governed by their variation. Instead, as mentioned above, the fluid dynamics are governed by variation of the gravitational degrees of freedom via the Bianchi identities. This is in-line with the usual treatment of fluids in classical general relativity where one does not specify the fluid equations of motion separately. 

Next, all 4-dimensional quantities need to be split into a 3+1 decomposition, and the 4-metric components replaced by functions of $N$, $N^{a}$ and $q^{ab}(E_{i}^{\;c})$. For this we write $[u_{\mu}]=[u,\, u_{a}]$ and $[s_{\mu}]=[s,\, s_{a}]$. After some calculation, the matter action in 3+1 form in the appropriate variables is found to be:
\begin{align}
I_{\af}=&8\pi\int_{M^{4}} N\sqrt{\detE}\left\{ \Theta \left[2uu_{b}\frac{N^{b}}{N^{2}} +\left(\qe{i}{a}{i}{b} -\frac{N^{a}N^{b}}{N^{2}}\right)u_{a}u_{b} -\frac{u^{2}}{N^{2}}+1\right] \right. \nonumber \\
&\qquad\qquad-2\pperp +\Delta \left.\left[2ss_{b}\frac{N^{b}}{N^{2}} +\left(\qe{i}{a}{i}{b} -\frac{N^{a}N^{b}}{N^{2}}\right)s_{a}s_{b} -\frac{s^{2}}{N^{2}}-1\right]\right\}d^{3}\mathbf{x}\,dt\,. \label{eq:decompmatteract}
\end{align}

Having now established a matter field in the correct variables, we freeze the symmetry of the system to spherical symmetry. For studies in the Ashtekar variables, we require an ansatz for a connection,  $A=A^{i}_{\;a}\,\tau_{i}dx^{a}\,$, as well as a densitized triad, $E=E_{i}^{\;a}\,\tau^{i}\partial_{a}\,$, which is capable of accommodating this symmetry. We therefore utilize the following pair for this:
\begin{subequations}
\romansubs
 \begin{align}
A=&\athree \tau_{1}\,dy + \left(\aone \tau_{2} + \atwo \tau_{3}\right)d\theta +\left(\aone\tau_{3} -\atwo\tau_{2}\right)\sin(\theta)\,d\phi +\cos(\theta)\tau_{1}\,d\phi\,, \label{eq:sphereA} \\
E=& \ethree \tau_{1} \sin(\theta) \frac{\partial}{\partial y} + \left(\eone\tau_{2} +\etwo\tau_{3}\right)\sin(\theta) \frac{\partial}{\partial \theta} +\left(\eone\tau_{3}-\etwo\tau_{2}\right) \frac{\partial}{\partial\phi}\,, \label{eq:sphereE} 
 \end{align}
\end{subequations}
with $0 < \phi \leq 2\pi$, $0 < \theta < \pi$ and the functions $\mathcal{A}_{..}$ and $\mathcal{E}_{..}$ are functions of the inhomogeneous coordinate, $y$ and time only. The $\tau_{i}$ represent the standard su(2) generators.

By utilizing (\ref{eq:sphereA}) and (\ref{eq:sphereE}) in (\ref{eq:vecconst}) and (\ref{eq:gaussconst}), and adding to them the matter contributions (from the variations of (10) with respect to the shift-vector for the vector constraint) we arrive at the following pair of constraints for our system:
\begin{subequations}
\romansubs
\begin{align}
V=&2\sin(\theta)\left\{\athree\left[\eone\atwo-\etwo\aone\right]-\eone\aone^{\prime} -\etwo\atwo^{\prime}\right\} \nonumber \\
 &+16\pi \sqrt{\left|\ethree\right|} \cdot \left|\sin(\theta)\right|\cdot \mathsf{E}\cdot \left[\Theta \, {U \cdot u_{x}} + \Delta \cdot {S\cdot s_{x}}\right]=0\,, \label{eq:spherevec}\\[0.1cm]
G=&\sin(\theta) \left[2\aone\etwo-2\atwo\eone +\ethree^{\prime}\right]=0\,, \label{eq:spheregauss} 
\end{align}
\end{subequations}
where primes denote derivatives with respect to the inhomogeneous coordinate $y$, and we have dropped the index on $V$ and $G$ as each constraint only yields one equation. Here we have used the definitions $\mathsf{E}:=\sqrt{\eone^{2} + \etwo^{2}}$, $U:=u/N^{2}$ and $S:=s/N^{2}$ to simplify expressions. Also, we have set the shift vector to zero after variation as there is no loss of generality by doing this since we are restricting ourselves to spherical symmetry. Note that the fluid does not contribute to the Gauss constraint. Also of note is that in the case of a static fluid, the fluid makes no contribution to the vector constraint\footnote{It may be of interest to note that the staticity of the resulting space-time is actually encoded in this statement, although it is not obvious. Time derivatives of the metric are encoded in the vector constraint via the connection $A^{i}_{\;a}$ through the extrinsic curvature in (\ref{eq:biconnection}). If the matter contribution to the vector constraint vanishes, the only way to satisfy the spherically symmetric  gravitational vector constraint is to make these terms involving the time derivative of the metric equal to zero.}.

For the scalar constraint the issue is more complicated. This is because in the gravitational part of $S$, (\ref{eq:scaleconst}), the extrinsic curvature terms appear and the extrinsic curvature needs to be replaced with the connection $A$ via (\ref{eq:biconnection}). Therefore we write
\begin{equation}
 K^{i}_{\;a}=\frac{1}{\gamma}\left[A^{i}_{\;a} - \Gamma^{i}_{\;a}\right]\,. \label{eq:kdef}
\end{equation}
However, $\Gamma^{i}_{\;a}$ is a function of the triad and inverse triad via (\ref{eq:Gam}), and this must be expressed in terms of the densitized triad $E_{i}^{\;a}$ only. Therefore we need to utilize the following relationships in (\ref{eq:Gam}):
\begin{equation}
 e_{i}^{\;a}=\frac{1}{\sqrt{\mbox{det}(E)}}\,E_{i}^{\;a}\,, \;\;\left[e^{i}_{\;a}\right]=\left[e_{i}^{\;a}\right]^{-1}=\sqrt{\mbox{det}(E)}\,\left[E_{i}^{\;a}\right]^{-1}\,. \label{eq:eofE}
\end{equation}
This yields the following spin-connection:
\begin{align}
\Gamma=&\frac{1}{\mathsf{E}^{2}}\left[\etwo\eone^{\prime}-\eone\etwo^{\prime}\right]\tau_{1}\,dx +\frac{\ethree^{\prime}}{2\mathsf{E}^{2}}\left[\eone\tau_{3}- \etwo\tau_{2}\right]d\theta -\frac{\ethree^{\prime}\sin(\theta)}{2\mathsf{E}^{2}} \left[\eone\tau_{2}+\etwo\tau_{3}\right]d\phi \nonumber \\
& +\cos(\theta)\tau_{1}d\phi\,.
\end{align}
 
We now have all the quantities required to write the full scalar constraint in these variables as
\begin{align}
\mathcal{S}=& \frac{2 |\sin(\theta)|}{\sqrt{\ethree}\cdot \mathsf{E}} \Big[2\eone\ethree\atwo^{\prime} -2 \etwo\ethree\aone^{\prime} +2\ethree\athree\left(\eone\aone+\etwo\atwo\right)+\mathsf{E}^{2}\left(\mathsf{A}^{2}-1\right)\Big] \nonumber \\
&-\frac{(1+\gamma^{2})|\sin(\theta)|}{2\sqrt{\ethree}\cdot \mathsf{E}^{3}\gamma^{2}} \Big[4\mathsf{E}^{4}\mathsf{A}^{2} +\mathsf{E}^{2} \left(\ethree^{\prime}\right)^{2} +4\mathsf{E}^{2} \ethree^{\prime} \left(\etwo\aone-\eone\atwo\right)  \nonumber \\
&\qquad\qquad\qquad\qquad\quad+8\ethree\left(\mathsf{E}^{2}\athree+\etwo^{\prime}\eone-\eone^{\prime}\etwo\right)\left(\etwo\atwo+\eone\aone \right)\Big] \nonumber \\
& +16\pi \sqrt{\left|\ethree\right|} \cdot \left|\sin(\theta)\right|\cdot \mathsf{E}\cdot \left[\Theta \, U\cdot u+ \Delta\cdot S\cdot s -\pperp\right]\,, \label{eq:bigscalar}
\end{align}
where the definition $\mathsf{A}:=\sqrt{\aone^{2}+\atwo^{2}}\,$, and the fluid constraints (\ref{eq:fluidvecs}), have been used after variation.

For the effective quantization we utilize a scheme that has proved fruitful in studies of loop quantum cosmology \cite{ref:bojlrr} and black holes \cite{ref:modesto}, \cite{ref:modesto2}, \cite{ref:BandV}. That is, we replace the connection with functions that reflect the holonomy structure of the representation of the quantum algebra. The operator $\widehat{A}$ is constructed from  turning the Poisson brackets to commutators:
\begin{equation}
 \left[\widehat{A}^{i}_{\;a}(\mathbf{x}),\,\widehat{E}_{j}^{\;b}(\mathbf{y})\right]=i\hbar \delta^{b}_{a} \delta^{i}_{j}\,\delta^{3}(\mathbf{x},\,\mathbf{y})\,. \label{eq:aecommutator}
\end{equation}
Although $\widehat{A}^{i}_{\;a}$ represented as ``multiplication by ${A}^{i}_{\;a}$'' can be used to satisfy the above bracket, it turns out that the operator $\widehat{A}^{i}_{\;a}$ does not have a well defined representation in the Hilbert space of connections. However, the space of functions of holonomies (cylindrical functions) does have a well defined, metric independent measure and can be turned into a mathematically rigorous Hilbert space. The holonomy of $A$ is given by
\begin{equation}
 h_{e}(A)=\mathcal{P}\,\exp\left[\int_{e} A\right]\,, \label{eq:holonomydef}
\end{equation}
where $\mathcal{P}$ represents a path ordering of the holonomy paths, $e$. In this vein we make the substitution,
\begin{equation}
 \mathcal{A}_{J} \rightarrow \frac{\sin\left(\mathcal{A}_{J}\,\delta_{J}\right)}{\delta_{J}}\,, \label{eq:sinrepgen}
\end{equation}
where $\delta_{J}$ represents the length of the holonomy path. This holonomy path is not constant, but related to the proper area of the holonomy loop, as has been shown in studies of loop quantum cosmology and black holes \cite{ref:BandV}, \cite{ref:APS}, \cite{ref:chiou}. More details of this will be provided later. Admittedly, this method is more rigorously motivated in certain studies of loop quantum cosmology \cite{ref:chiou}, \cite{ref:sinreppaper}.

Although not required for the study of the constraints, for completion we construct the fluid Hamiltonian by identifying appropriate canonical momenta for the fluid:
\begin{equation}
\PI{1}{\mu}:=-16\pi N\sqrt{\detE} \cdot \Theta \cdot u_{\mu}\,,\;\;\; \PI{2}{\mu}:=-16\pi N\sqrt{\detE} \cdot \Delta \cdot s_{\mu}\,, \label{eq:fluidmoms}
\end{equation}
which leads, via the standard Legendre transformation, to the matter Hamiltonian
\begin{equation}
 H_{\af}=\int \mathcal{H}_{\af}\,d^{3}\mathbf{x}\:=\: 16\pi\int N \sqrtdetE \cdot \left[\Theta - \Delta - \pperp\right]\,d^{3}\mathbf{x}\,. \label{eq:fluidham}
\end{equation}
(Care must be taken with minus signs as our Lagrangians are ``minus'' the more common convention.) In the case of the perfect fluid, the above reduces simply to an integral over the proper-energy density of the fluid.

The above analysis is general, save for the assumption of spherical symmetry imposed from equations (\ref{eq:sphereA}) and (\ref{eq:sphereE}) onward, and rather complex. We next simplify the system to study specific cases of interest. 
\vspace{0.75cm}

{\centering\subsection{\hspace{-0.35cm}:\hspace{0.33cm}The static scenarios}}
The models we shall study here will be time independent. Under the added assumption of staticity, the vector constraint reduces simply to the gravitational vector constraint, as a static fluid makes no contribution to the vector constraint. We shall also impose the following simplification on the connection and densitized triad, which is still compatible with spherical symmetry at least in the static case:
\begin{subequations}
\romansubs
\begin{align}
 A=&  \atwo \tau_{3}\,d\theta - \atwo\tau_{2}\sin(\theta)\,d\phi +\cos(\theta)\tau_{1}\,d\phi\,, \label{eq:simpA} \\
 E=& \ethree \tau_{1} \sin(\theta) \frac{\partial}{\partial y} + \eone\tau_{2} \sin(\theta) \frac{\partial}{\partial \theta} +\eone\tau_{3} \frac{\partial}{\partial\phi}\,, \label{eq:simpE}
\end{align}
\end{subequations}
where now the functions $\mathcal{A}$ and $\mathcal{E}$ are functions of $y$ only. The relationship between the functions above and the usual metric in curvature coordinates is the following (with $y$ in this case identified as the usual radial coordinate $r$) from the relation $\mbox{det}(q)\cdot q^{ab}=E_{i}^{\;a} \cdot E_{i}^{\;b}\:$:
\begin{align}
 d\sigma^{2}=q_{ab}\,dx^{a}\,dx^{b}= &B(r)\,dr^{2}+r^{2}\,d\theta^{2} +r^{2}\sin^{2}\theta\,d\phi^{2} \nonumber \\[0.2cm] =&\frac{(\eone)^{2}}{\ethree}\,dr^{2}+\ethree\,d\theta^{2} +\ethree\sin^{2}\theta\,d\phi^{2}\,.
\end{align}
From this choice, the remaining gravitational vector constraint is identically satisfied and the Gauss constraint (\ref{eq:spheregauss}) is satisfied by employing the following relation:
\begin{equation}
 \ethree^{\prime}=2\atwo\eone\,. \label{eq:earel}
\end{equation}

At this stage it is worth noting a few issues regarding the symplectic structure. By making the above simplification we are merely going to another set of canonical coordinates. To see this, one can make the following transformation on the original set of variables
\begin{equation}
 \aone=:\aphi\cos(\beta)\,, \quad \atwo=:\aphi\sin(\beta)\,, \label{eq:betaatransf}
\end{equation}
for some angle $\beta$. Our particular choice corresponds to $\beta=\pi/2$. Similarly, via rotations on the densitized triad (both spatial and SU(2)), a canonical momentum can be defined via
\begin{equation}
 \pi^{\phi}:= 2\eone\ =: 2\ephi\,. \label{eq:betaetransf}
\end{equation}
These configuration-momentum coordinates turn out to be exactly those of Bojowald and Swiderski \cite{ref:BojSwi}, and Campiglia, Gambini and Pullin \cite{ref:campschwarz}, \cite{ref:gambpul} with symplectic structure:
\begin{equation}
 \left\{\aphi(\mathbf{x}),\,\pi^{\phi}(\mathbf{x^{\prime}})\right\} =2\gamma \delta(\mathbf{x}-\mathbf{x}^{\prime})\,. \label{eq:symstruc}
\end{equation}
Below we shall continue using the notation $\atwo$ and $\eone$ for these components.

Using the above simplification yields only a single connection component, $\atwo$, that needs to be effectively quantized. Using the method discussed earlier, we make the substitution
\begin{equation}
 \atwo \rightarrow \frac{\sin\left(\atwo\,\deltatwo\right)}{\deltatwo}\,, \label{eq:sinrep}
\end{equation}
where again $\deltatwo$ represents the length of the holonomy path. As mentioned previously, it has been shown in a cosmological setting in \cite{ref:APS} \cite{ref:chiou} and for black hole interiors in \cite{ref:BandV}, that a reasonable semi-classical limit might not be obtained utilizing a fixed $\delta$ in the above substitution. Instead, one can relate the length of the holonomy paths (the $\delta$'s) to the classical area. The proper area spanned by a holonomy ``square-loop'' in the $\theta-\phi$ sector is given by:
\begin{equation}
 a_{\mbox{\tiny{$\theta\phi$}}} \approx\ethree  \left(\deltatwo\right)^{2}\,. \label{eq:adeltarel}
\end{equation}
By setting the above area equal to the smallest area predicted by the spectrum of the area operator of loop quantum gravity, we arrive at what is known as the $\overline{\mu}^{\prime}$ scheme\footnote{That is, the ``mu-bar-prime'' scheme. The symbol $\mu$ is often used instead of $\delta$ for the length of the holonomy path in loop quantum cosmology.} \cite{ref:APS}. In full loop quantum gravity, the infinitesimal area operator is given by
\begin{equation}
 \widehat{da}=\left[n_{a}\widehat{E}_{k}^{\;a}\,n_{b}\widehat{E}_{k}^{\;b}\right]^{1/2}\,, \label{eq:area}
\end{equation}
and has the following spectrum:
\begin{equation}
 \widehat{da}\left|\mathbf{S}\right> =8\pi\gamma \ell^{2}_{\mbox{\tiny{p}}} \sqrt{j_{p}(j_{p}+1)}
 \left|\mathbf{S}\right>, \label{eq:aevals}
\end{equation}
with $\ell^{2}_{\mbox{\tiny{p}}}$ the Planck length. The integer $p$ denotes which puncture (or area element) is
under consideration, and $j_{p}$ can take on half-integer values which represent spins carried by the punctures.
The covariant normal vectors to the surface element are denoted by $n_{a}$ and $n_{b}$. Setting $j_{p}$ equal to one-half yields the smallest predicted area, which we denote as $\mina$:
\begin{equation}
 \mina:=\sqrt{3}\,4\pi\gamma \ell^{2}_{\mbox{\tiny{p}}}\,, \label{eq:Delta}
\end{equation}
so that, by (\ref{eq:adeltarel}), we have $\deltatwo= \sqrt{\mina}/\sqrt{\ethree }\,$. The results presented below are valid for any value of $\mina$, provided it is small. The Immirzi parameter, $\gamma$, is determined by some means such as black hole entropy calculations \cite{ref:entrev2}- \cite{ref:corichioverview}. Putting everything together, the effective quantization amounts to the replacement
\begin{equation}
 \atwo \rightarrow \frac{\sin\left(\atwo\,\frac{\sqrt{\mina}}{\sqrt{\ethree}}\right) \cdot \sqrt{\ethree}}{\sqrt{\mina}}\,, \label{eq:deltasinrep}
\end{equation}
with $\mina$ provided by (\ref{eq:Delta}).

At this stage, the only constraint which remains to be satisfied is the scalar constraint (\ref{eq:bigscalar}). With the imposition of the other constraints and staticity, the scalar constraint simplifies to
\begin{align}
 \mathcal{S}=& \frac{2|\sin(\theta)|}{\sqrt{\ethree} |\eone|} \left[2\eone\ethree\atwo^{\prime} +\eone^{2}\left(\atwo^{2}-1\right)\right] -\frac{(1+\gamma^{2})|\sin(\theta)|}{2\sqrt{\ethree}|\eone|^{3}\gamma^{2}} \left[4\eone^{4}\atwo^{2}+\eone^{2} (\ethree^{\prime})^{2} -4\eone^{3}\ethree^{\prime}\atwo\right] \nonumber \\[0.2cm]
&+16\pi\sqrt{\left|\ethree\right|} \cdot \left|\sin(\theta)\right|\cdot |\eone| \cdot \left[\Theta \, U\cdot u -\pperp\right] \,. \label{eq:reducedS}
\end{align}
Derivatives can be treated as usual continuous derivatives; specifically, for $\atwo$:
\begin{align}
& \atwo^{\prime}\rightarrow \left[\frac{\sin\left(\atwo\,\frac{\sqrt{\mina}}{\sqrt{|\ethree|}}\right) \cdot \sqrt{|\ethree|}}{\sqrt{\mina}}\right]^{\prime} \nonumber \\[0.2cm]  
&\quad  = \frac{\ethree^{\prime}}{2\sqrt{|\ethree|}\sqrt{\mina}} \sin\left(\frac{\atwo\,\sqrt{\mina}}{\sqrt{|\ethree|}}\right)
+\cos\left(\frac{\atwo\,\sqrt{\mina}}{\sqrt{|\ethree|}}\right) \left[\atwo^{\prime} -\frac{\atwo}{2}\frac{\ethree^{\prime}}{\ethree}\right]. \label{eq:Aderiv}
\end{align}
Alternatively, one can initially treat the derivative as a finite-difference, as suggested in \cite{ref:gambpul} for vacuum black holes:
\begin{align}
&\atwo^{\prime}\rightarrow \nonumber \\
&\frac{1}{\epsilon}\left\{\left[\frac{\sin\left(\frac{\atwo\,\sqrt{\mina}}{\sqrt{\ethree}}\right) \sqrt{|\ethree|}}{\sqrt{\mina}}\right]_{x+\epsilon} - \left[\frac{\sin\left(\frac{\atwo\,\sqrt{\mina}}{\sqrt{|\ethree|}}\right) \sqrt{\ethree}}{\sqrt{\mina}}\right]_{x}\right\}\,,
\end{align}
where we use a constant discretization, $\epsilon$\,, for the derivative as the derivative is taken with respect to \emph{coordinate distance}, not a holonomy path.

\begin{figure}[h!t]
\begin{framed}
\begin{center}
\vspace{0.5cm}
\includegraphics[bb=0 0 544 435, scale=0.4, clip, keepaspectratio=true]
{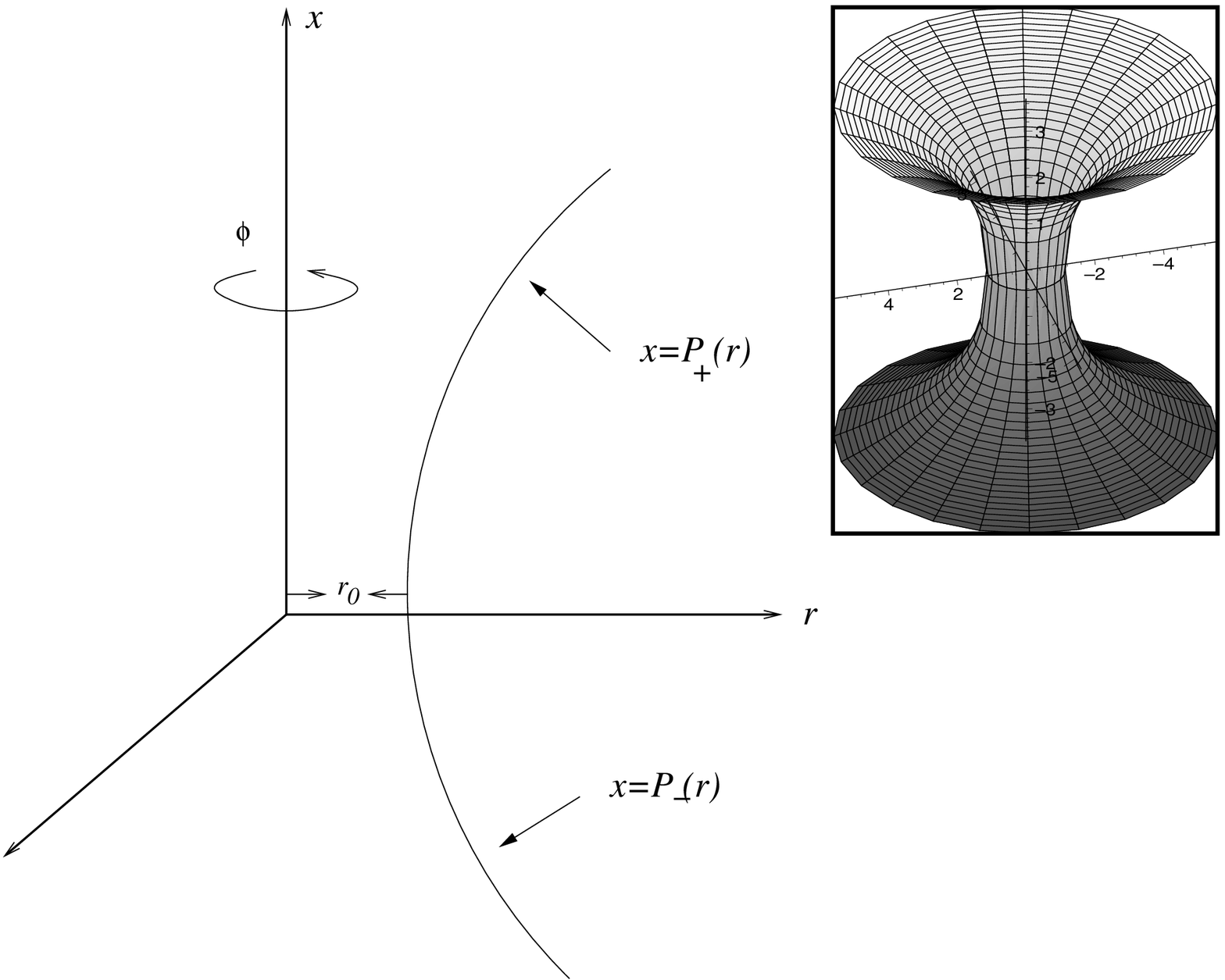}
\end{center}\vspace{-0.1cm}
\caption{{Wormhole profile curve, $P_{\pm}(r)$, in the $\theta=\pi/2$ submanifold. The wormhole is generated via
rotation about the $x$-axis (inset).}}%
\label{fig:curve}
\end{framed}
\end{figure}

\vspace{0.7cm}
{\centering\subsubsection{\normalsize\hspace{-0.35cm}:\hspace{0.23cm} Wormhole}}
The above analysis in usual curvature coordinates is suitable for many spherically-symmetric anisotropic structures. However, as mentioned in the introduction, of particular importance in studies of quantum gravity is the wormhole. In curvature coordinates a sub-manifold of the wormhole is generated by considering a curve with the correct properties, and then creating a surface of revolution from this profile curve. A sample profile curve, along with the corresponding surface of revolution is illustrated in figure~\ref{fig:curve}. Note that in this chart two profile curves are required, one for the ``top-part'' of the wormhole ($P_{+}(r)$) and one for the ``bottom-part'' of the wormhole ($P_{-}(r)$).

\begin{figure}[h!t]
\begin{framed}
\begin{center}
\vspace{0.5cm}
\includegraphics[bb=0 0 660 410, scale=0.4, clip, keepaspectratio=true]
{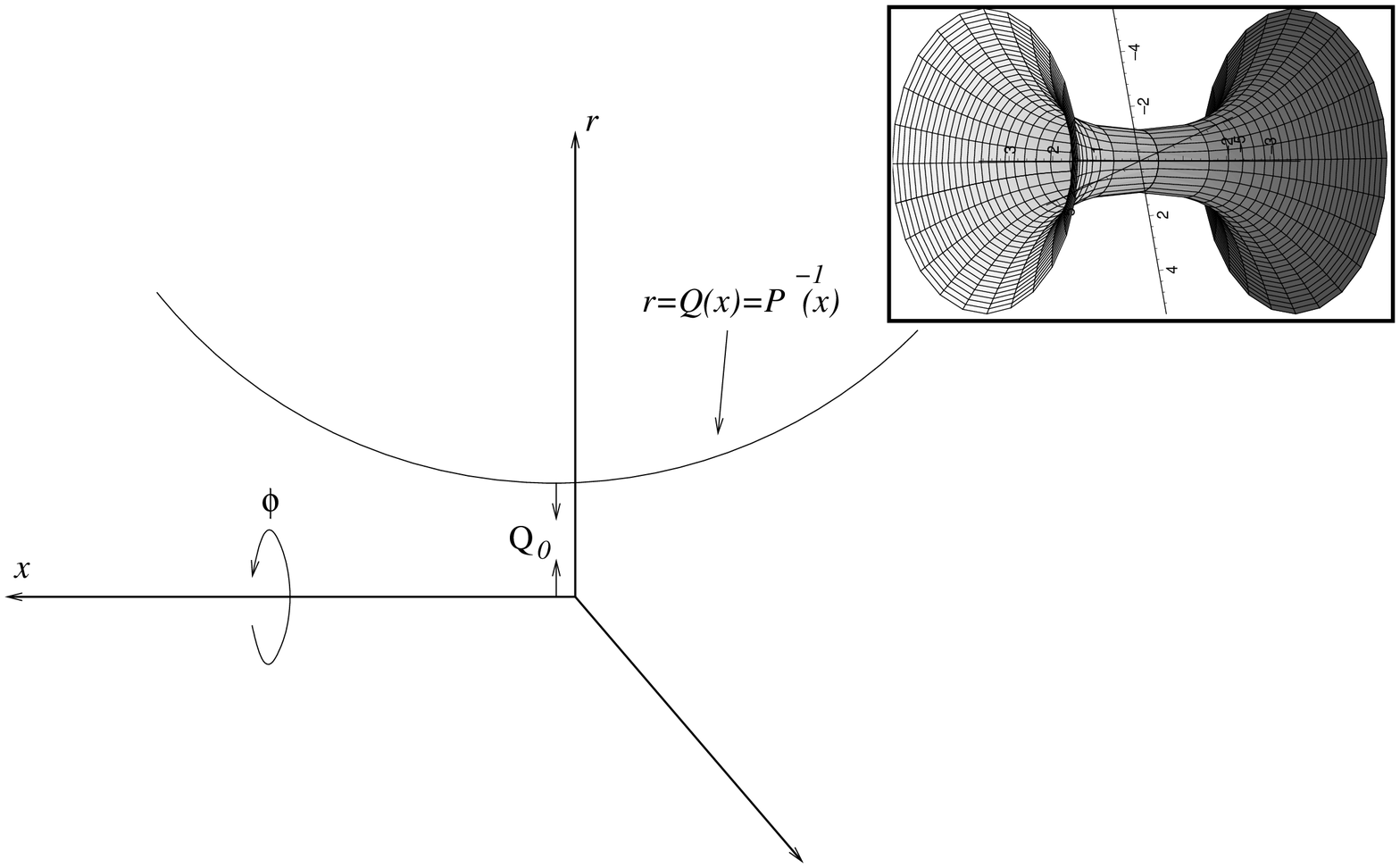}
\end{center}\vspace{-0.1cm}
\caption{{Wormhole profile curve in the $\theta=\pi/2$ submanifold using the rotated coordinate system. The profile
function is given by $r=Q(x)=P^{-1}(x)$ and the radius of the throat is $Q_{0}$. As before, the wormhole is generated via rotation about the $x$-axis (inset). Only a single profile function $Q(x)$ is needed now and the densitized triad components are finite at the throat radius.}}%
\label{fig:rotated}
\end{framed}
\end{figure}

In terms of the profile curve function, $P_{\pm}(r)$, the densitized triad components read:
\begin{equation}
\eone=r\sqrt{B(r)}=r\left\{\grr\right\}^{1/2},\quad \ethree=r^{2}\,, \label{eq:eofp}
\end{equation}
with $r \geq r_{0}$ (refer to figure). Immediately one can see a potential problem with the variables chosen when discussing the wormhole scenario. Note that at the wormhole throat ($r_{0}$ in figure~\ref{fig:curve}) the derivative of $P(r)$ becomes infinite, and hence the densitized triad component $\eone$ is badly behaved at the throat. This malady affects the metric formulation as well. Hence, an alternative to the usual curvature coordinates will be used here. We use a system of coordinates introduced in \cite{ref:debdasworm}, and utilized in \cite{ref:debdashighworm} and \cite{ref:fureydeb}, which is better behaved at the throat. Essentially, we rotate the entire chart by $\pi/2$ as shown in figure~\ref{fig:rotated}. 

In this new coordinate system, the spherically symmetric anzats (\ref{eq:sphereA}) and (\ref{eq:sphereE}) still hold and the equations in this chart are only slightly more complicated than in the usual curvature coordinates. The line element is now
\begin{align}
 d\sigma^{2}=q_{ab}\,dx^{a}\,dx^{b}= &C(x)\,dx^{2}+Q^{2}(x)\,d\theta^{2} +Q^{2}(x)\,\sin^{2}\theta\,d\phi^{2} \nonumber \\[0.2cm] =&\frac{(\eone)^{2}}{\ethree}\,dx^{2}+\ethree\,d\theta^{2} +\ethree\sin^{2}\theta\,d\phi^{2}\,. \label{eq:spacemetQ}
\end{align}
Here we have used the same simplification for $A$ and $E$ as we used previously. Note that the advantage of this chart is that the relation
\begin{equation}
 \eone=Q(x)\sqrt{C(x)}=Q(x)\left\{\gxx\right\}^{1/2},\quad \ethree=\left[Q(x)\right]^{2}\,. \label{eq:eofq}
\end{equation}
yields finite components at the throat as the derivative of $Q(x)$ \emph{is zero, not infinite} there. As well, the coordinate $x$ may span $-\infty < x < \infty$ and hence it is now possible to cover the wormhole geometry with just a single well-behaved chart (if the poles of the two-spheres, $\theta=0$ and $\theta=\pi$, are excluded).

Again in the static case, the vector constraint is satisfied. The Gauss constraint is also again satisfied by imposing the relation (\ref{eq:earel}). However, this condition is slightly more complicated in this chart due to the fact that the derivative of $\ethree$ contains the derivative of $Q(x)$.

We do not present all the details here as the procedure is similar to the above. In short, the effective quantization is accomplished by the same substitution as previously done:
\begin{equation}
 \atwo \rightarrow \frac{\sin\left(\atwo\,\frac{\sqrt{\mina}}{\sqrt{\ethree}}\right) \cdot \sqrt{\ethree}}{\sqrt{\mina}}\,, \label{eq:deltasinrep2}
\end{equation}
and the simplified scalar constraint reads as before (\ref{eq:reducedS}).

The most salient features of a wormhole occur in the throat region. For example, in classical wormholes, the necessary violation of energy conditions in static wormholes occurs in the neighborhood of the throat. In fact, in the classical scenarios if the matter field falls off sufficiently fast, or is patched to a vacuum solution, the properties far from the throat do not differ greatly from similar systems with trivial topology. As well, it has been argued in \cite{ref:hochvis} that the global topology is too limited a tool to study wormholes, and a local geometric analysis near the throat is generally more useful in discerning interesting properties of wormholes. We therefore now concentrate on the near throat region and study the quantum gravity corrections in this vicinity.

In the neighborhood of the throat ($x=0$), the profile function, $Q(x)$ must have the following properties:
\begin{enumerate}[i)]
\item $\qzero:=Q(0) > 0$,\vspace{-0.2cm}
\item $Q^{\prime}(x)_{|x=0}=0$,\vspace{-0.2cm}
\item $Q^{\prime\prime}(x) > 0$ in some neighborhood of the throat\footnote{More precisely, if $Q$'s first non-zero derivative (higher than first order) at $x=0$ is of even order, the function attains a local minimum if this derivative is positive, and hence we have a wormhole throat. If its first non-zero derivative is of odd order, it is a point of inflection and therefore does not describe a wormhole throat.}.
\end{enumerate}
Aside from the above properties we make the mild assumption that $Q(x)$ is analytic. Now, $\eone$ and $\ethree$ are functions of $Q(x)$ and its derivative via (\ref{eq:eofq}) and one can substitute these functions into (\ref{eq:reducedS}) to study the remaining scalar constraint. We shall first concentrate on the purely classical scenario, and hence simply replace $\atwo$ in (\ref{eq:reducedS}) by using (\ref{eq:earel}); i.e. $\atwo=\ethree^{\prime}/(2\eone)$. This yields the following scalar constraint:
\begin{equation}
 \mathcal{S}=\mathcal{S}_{\mbox{\tiny grav}}+\mathcal{S}_{\mbox{\tiny af}}=\frac{2|\sin(\theta)|}{\left[1+\left(Q^{\prime}\right)^{2}\right]^{\frac{3}{2}}} \left[2 Q \cdot Q^{\prime\prime} -\left(Q^{\prime}\right)^{2} -1\right] + 16\pi|\sin(\theta)|\cdot  Q^{2} \sqrt{1+\left(Q^{\prime}\right)^{2}}\: \rho \:.
\end{equation}
We treat $\rho$ as the unknown and hence the constraint $\mathcal{S}=0$ yields:
\begin{equation}
 8\pi\rho=\frac{\left[1-2 Q Q^{\prime\prime} + (Q^{\prime})^{2}\right]}{Q^{2}\left[1+ (Q^{\prime})^{2}\right]^{2}}\,.
\end{equation}
The above expression agrees with the expression one may derive by calculating the $t-t$ component of the Einstein tensor using metric (\ref{eq:spacemetQ}) and (\ref{eq:eofq}) using an arbitrary lapse (which does not appear in $G^{t}_{\;t}$). Next, the analyticity of $Q(x)$ allows for a Taylor expansion about $x=0$, resulting in the following near-throat expression:
\begin{align}
8\pi \rho=& \frac{1}{\qzero^{2}}\left(1-2\qzero^{\prime\prime}\qzero\right) - \frac{2\qzero^{\prime\prime\prime}}{\qzero}x -\frac{1}{\qzero^{3}}\left[ \qzero^{\prime\prime} + \qzero^{\prime\prime\prime\prime}\qzero^{2}  - 4 \left(\qzero^{\prime\prime}\right)^{3}\qzero^{2}\right] x^{2} \nonumber \\[0.2cm] 
& +\frac{1}{3\qzero^{3}} \left[\qzero^{\prime\prime\prime} \qzero^{\prime\prime} \qzero -\qzero^{\prime\prime\prime} - \qzero^{\prime\prime\prime\prime\prime}\qzero^{2} +24 \qzero^{\prime\prime\prime}\left(\qzero^{\prime\prime}\right)^{2} \qzero^{2}   \right]x^{3}    +\mathcal{O}(x^{4})\,, \label{eq:classicalexpansion}
\end{align}
where the subscript $0$ indicates that the quantity is evaluated at $x=0$. (We do not demand that terms odd in powers of $x$ vanish as the wormhole does not need to be symmetric about the throat.)

Next we tackle the more difficult case with the quantum holonomy corrections. In this case, $\atwo$ in (\ref{eq:reducedS}) is first replaced via (\ref{eq:deltasinrep2}), and the derivative of $\atwo$ is replaced by (\ref{eq:Aderiv}), after which the substitution $\atwo=\ethree^{\prime}/(2\eone)$ (from the Gauss constraint, (\ref{eq:earel})) is utilized. The full analytic expression for $\rho$ can be calculated but the result is quite complicated so we simply present the series result:
\begin{align}
8\pi\rho=& \frac{1}{\qzero^{2}}\left(1-2\qzero^{\prime\prime}\qzero\right) - \frac{2\qzero^{\prime\prime\prime}}{\qzero}x -\frac{1}{\qzero^{3}}\left[ \qzero^{\prime\prime} + \qzero^{\prime\prime\prime\prime}\qzero^{2}  - 4 \left(\qzero^{\prime\prime}\right)^{3}\qzero^{2} - \mina \left(\qzero^{\prime\prime}\right)^{3} \right] x^{2} \nonumber \\[0.2cm]
& +\frac{1}{3\qzero^{3}} \left[\qzero^{\prime\prime\prime} \qzero^{\prime\prime} \qzero -\qzero^{\prime\prime\prime} - \qzero^{\prime\prime\prime\prime\prime}\qzero^{2} +24 \qzero^{\prime\prime\prime}\left(\qzero^{\prime\prime}\right)^{2} \qzero^{2}  +6\mina \qzero^{\prime\prime\prime} \left(\qzero^{\prime\prime}\right)^{2} \right]x^{3}    +\mathcal{O}(x^{4})\,. \label{eq:quantumexpansion}
\end{align}
The purely classical limit is achieved when $\mina\rightarrow 0$, which agrees with (\ref{eq:classicalexpansion}). Note that the quantum corrections (terms multiplied by $\mina$) do not contribute exactly at the wormhole throat. Instead, they come in at $x^{2}$ order and, given the sign of the second derivative of $Q(x)$ near the throat, the quantum corrections contribute \emph{positively} to the energy density near the throat and hence may act to lessen energy condition violation for the matter in the vicinity of the throat. Even though the energy density can be positive at the throat even in the purely classical scenario, it is known that somewhere in the vicinity of the throat energy conditions must be violated (see, for example, \cite{ref:hochvis}). The fact that this modified theory of gravity increases the energy density is a positive indication for the lessening of energy condition violation, although the pressures should be also be studied for a concrete statement. We make some comments on the pressures below.

Another expansion, although perhaps not as useful, is an expansion of $\rho$ in powers of $\mina$. This yields an  expression that is order-by-order in powers of quantum corrections:
\begin{align}
8\pi\rho=&\frac{\left[1-2 Q Q^{\prime\prime} + (Q^{\prime})^{2}\right]}{Q^{2}\left[1+ (Q^{\prime})^{2}\right]^{2}} - \frac{ \left(Q^{\prime}\right)^{2}}{3 \left[1+(Q^{\prime})^{2}\right]^{3} Q^{4}} \Big[ (Q^{\prime})^{4} + (Q^{\prime})^{2} -3Q^{\prime\prime}Q\Big] \mina \nonumber \\[0.2cm]
&+\frac{(Q^{\prime})^{4}}{180\gamma^{2} \left[1+(Q^{\prime})^{2}\right]^{4} Q^{6}} \Big[ \left(9\gamma^{2} + 5\right) \left((Q^{\prime})^{4} + (Q^{\prime})^{2}\right) -15 \gamma^{2}Q^{\prime\prime}Q \Big] \mina^{2}  \nonumber \\[0.2cm]
&- \frac{(Q^{\prime})^{6}}{2520\gamma^{2} \left[1+(Q^{\prime})^{2}\right]^{5} Q^{8}} \Big[ \left(5\gamma^{2} + 7\right) \left((Q^{\prime})^{4} + (Q^{\prime})^{2}\right) -7 \gamma^{2}Q^{\prime\prime}Q \Big] \mina^{3} +\mathcal{O}(\mina^{4})\,. \label{eq:deltaexpansion}
\end{align}
Compatible with the previous result, it can be noted that at local extrema where $Q^{\prime}=0$ (eg. the throat), terms arising from the quantum holonomy corrections do not make a contribution, but do contribute slightly away from the extremal point. It is therefore possible in principle to have a model where the energy densities of both the classical and quantum corrected throat are zero at the throat, but in the vicinity of the throat region the energy density of the classical solution is negative and the quantum one is positive.

Ideally one would like to analyze the individual pressures to completely specify the properties of the matter field. However, in the 3+1 connection formalism this is not readily available\footnote{This would involve writing the 3+1 action (including the connection terms) completely in terms of the metric and its derivatives, vary the action with respect to the metric, then re-write terms which can be expressed as connections again as connection terms, so that the substitution (\ref{eq:deltasinrep}) may be performed to get the quantum corrected pressures.}. In the case of isotropic loop quantum cosmology with a scalar field, one can extract information on the (isotropic) pressure and this has been achieved in \cite{ref:coseconds} where energy conditions were studied.

\begin{figure}[h!t]
\begin{framed}
\begin{center}
\vspace{0.5cm}
\includegraphics[bb=0 0 1500 400, scale=0.29, clip, keepaspectratio=true]
{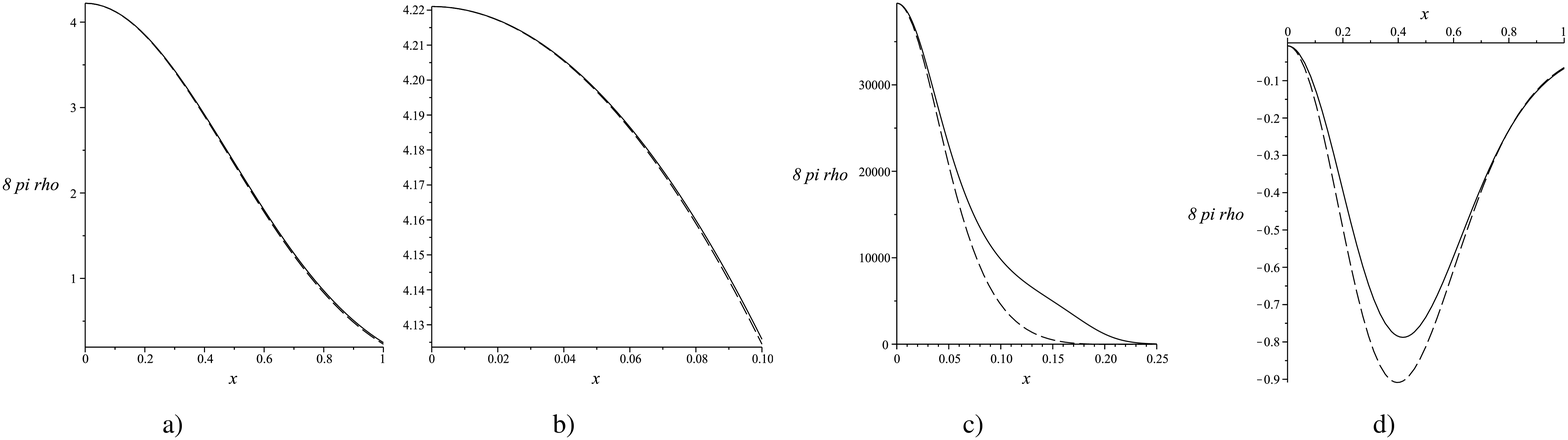}
\end{center}\vspace{-0.1cm}
\caption{{Symmetric wormhole models with $Q(x)=\alpha_{0}\cosh(x/x_{0})$. The parameters are as follows: Figure a): $\alpha_{0}=0.38,\,x_{0}=0.86$, figure b): A close up of the throat region of the previous figure, figure c): $\alpha_{0}=0.005,\,x_{0}=0.06$, d): $\alpha_{0}=0.3537,\,x_{0}=0.5$. In all cases $\gamma\approx0.27$ and $\ell_{\mbox{\tiny{p}}}$ was set to 0.1 to exaggerate the differences to make them easier to see.}}%
\label{fig:symmetricrho}%
\end{framed}
\end{figure}

\begin{figure}[h!t]
\begin{framed}
\begin{center}
\vspace{0.5cm}
\includegraphics[bb=0 0 800 450, scale=0.32, clip, keepaspectratio=true]
{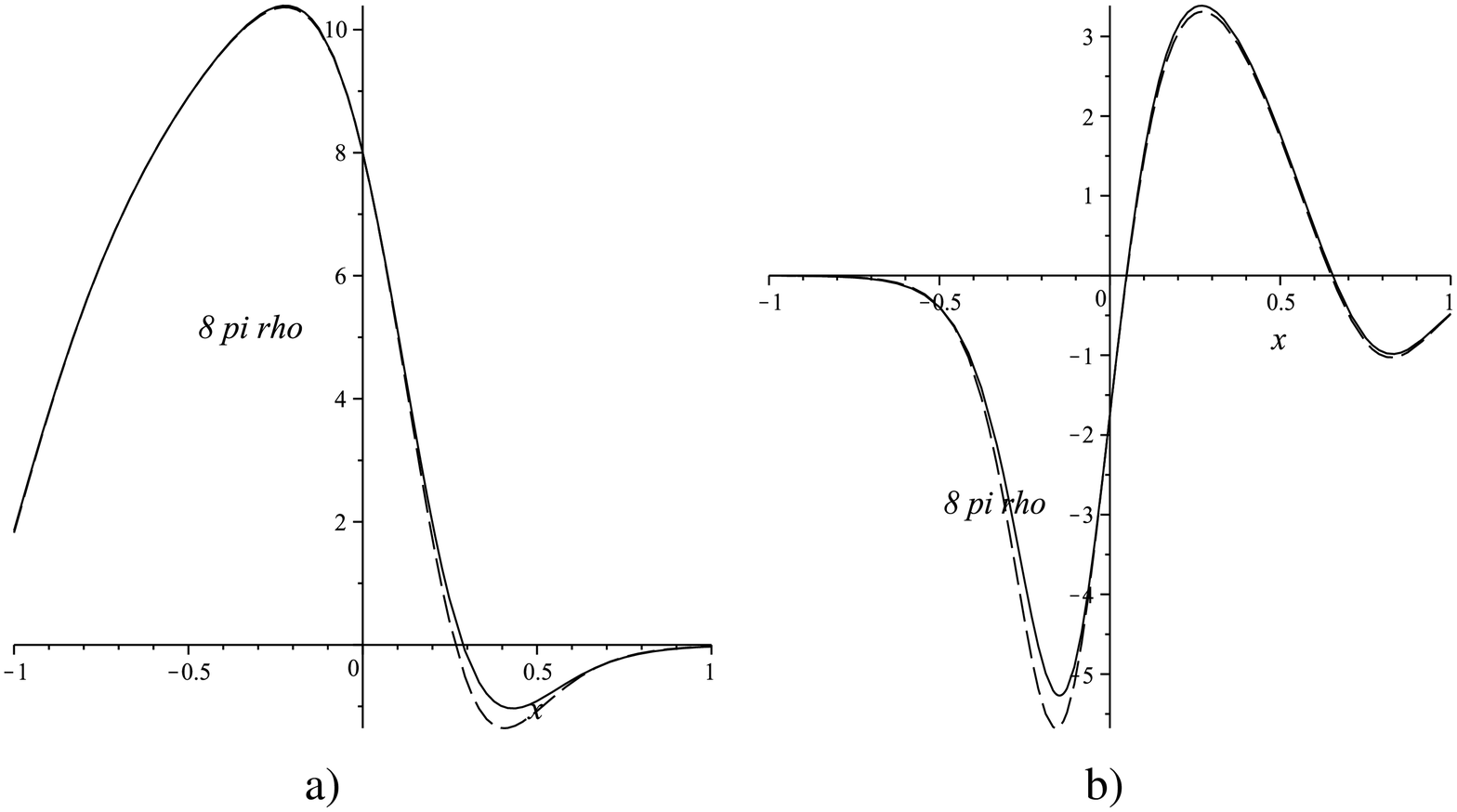}
\end{center}\vspace{-0.1cm}
\caption{{Non-symmetric wormhole models with $Q(x)=\alpha_{0}\cosh(x/x_{0})+\beta_{0}x^{3}$. The parameters are as follows: Figure a): $\alpha_{0}=0.25,\,x_{0}=0.5,\,\beta_{0}=0.5$, figure b): $\alpha_{0}=0.305,\,x_{0}=0.4,\,\beta_{0}=-0.95$. In both cases $\gamma\approx0.27$ and $\ell_{\mbox{\tiny{p}}}$ was set to 0.1 to exaggerate the differences to make them easier to see.}}%
\label{fig:nonsymmetricrho}%
\end{framed}
\end{figure}

Finally we present a few specific models to compare the quantum corrected models with the purely classical models. In figures \ref{fig:symmetricrho}a-d) a specific wormhole profile $Q(x)=\alpha_{0}\cosh(x/x_{0})$ is chosen and the quantum corrected (solid lines) and classical (dashed) energy densities are plotted for various values of the parameters, with both positive and negative energy densities near the throat. Note that the quantum effects tend to raise the energy of the fluid, and hence the exotic nature of the fluid tends to be lessened in comparison to the purely classical scenarios in the cases where the energy densities are negative. In figures \ref{fig:nonsymmetricrho}a-b) we also plot non-symmetric wormholes with a profile given by $Q(x)=\alpha_{0}\cosh(x/x_{0})+\beta_{0}x^{3}$ for sufficiently small $\beta_{0}$ so as not to spoil the local minimum. It can also be seen here that the quantum corrected versions tend to have more positive energy densities. In all cases where negative energy densities occur, the region of negative energy is smaller in the corresponding quantum corrected case. It is possible that those particular models which have outwardly increasing energy density may be unstable. However, if these models are to represent ephemeral quantum fluctuations in the vacuum, it may be that instability is not a serious issue. 

In classical general relativity, there are singularity theorems which state that if the classical equations of motion hold in scenarios describing objects such as gravitational collapse to black holes or collapsing cosmologies, and the material present obeys energy conditions, then a singularity is inevitable \cite{ref:singthm1}, \cite{ref:singthm2}. One loop-hole out of these arguments is the abandoning of the energy conditions, as it is known that energy condition violating matter can theoretically prevent such singularities from forming. It is also often believed that quantum gravity effects should take over close to singularity formation, when the curvature is large, and that these effects may prevent the formation of the classical singularity. In this vein, the loop quantum gravity paradigm has previously indicated that both in cosmology and black holes the quantum gravity effects have replaced the singularity with a smooth bounce \cite{ref:ashbounce}, \cite{ref:bojolrr}, \cite{ref:modesto}, \cite{ref:BandV} due to gravity becoming repulsive under certain situations in the quantum regime. That is, a scenario that classically requires matter energy condition violation to occur, can occur naturally within loop quantum gravity. What we find here in the context of wormholes is a manifestation of the same phenomenon. In order to support the wormhole classically, energy conditions must be violated to a certain degree. In the quantum corrected case, however, the energy condition violation is less, and this is directly due to the quantum properties of the gravitational field. It is, in principle from the above expressions, possible to find wormhole solutions where there is an energy density which is everywhere positive (with a zero at the throat) but whose classical counter-part yields an everywhere negative (with a zero at the throat) fluid energy density. This is due the  repulsive nature of gravity in the regime where loop quantum effects become important. It should be noted that these are low-order quantum corrections, similar to those employed in most studies of effective loop quantum black holes or loop quantum cosmology. It is a possibility that higher order corrections may improve the energy condition situation even further. Related to this, the repulsive nature of the gravitation under such extreme conditions could act to alleviate the singularities that are thought to occur when space-time topology changes classically. (Although even classically it has been shown that there are some topology changing scenarios where the singularities are not curvature singularities but merely metric ones \cite{ref:horo}.)  

\vspace{0.4cm}
\begin{center}\section{\hspace{-0.35cm}:\hspace{0.33cm}\small CONCLUDING REMARKS}\end{center}
We have studied anisotropic spherically symmetric systems in the 3+1 Hamiltonian formalism of gravity in the variables used in loop quantum gravity. By replacing the SU(2) connection with functions that encode the holonomy structure of the corresponding operator we essentially have an effective theory with some quantum inspired corrections. In the case of wormhole throats it was found that the energy density of the material source is increased in comparison to the purely classical case. This may indicate that the energy condition violation ubiquitous in wormhole throats in Einstein gravity may be lessened by quantum gravity effects. This provides another arena to study higher energy effects of gravity, complementing other techniques such as those incorporating higher curvature effects through a modified gravitational Lagrangian (such as $f(R)$ theories \cite{ref:fureydeb}, \cite{ref:fofr1}-\cite{ref:fofrlast}). As well, the wormhole provides yet another theater to test the predictions of a quantum gravity theory.

\vspace{0.65cm}
{\centering\section*{\small ACKNOWLEDGMENTS}\vspace{-0.1cm}}
The author is grateful for discussions with F.S.N. Lobo, W.-T. Ni, F. Rahaman, S. Ray and A. Usmani. The author would also like to acknowledge the kind hospitality of the Inter-University Centre for Astronomy and Astrophysics, Pune, where some of this work was carried out.
\vspace{0.55cm}

\linespread{0.75}
\bibliographystyle{unsrt}

\end{document}